\documentclass[aps,pra,reprint,twocolumn,superscriptaddress,showpacs,showkeys,superscriptaddress,longbibliography]{revtex4-1}
\usepackage{amsmath,amssymb,amstext}
\usepackage[usenames,dvipsnames]{color}
\usepackage{graphicx}
\usepackage{bm,bbold,braket}
\usepackage{natbib}
\usepackage{txfonts, comment,stmaryrd}
\usepackage{dcolumn}
\usepackage{color}
\usepackage{xcolor}
\colorlet{rn}{red}
\colorlet{an}{blue}
\usepackage[english]{babel}

\usepackage[colorlinks,bookmarks=false,citecolor=blue,linkcolor=red,urlcolor=blue]{hyperref}
\begin{document}

\title{Population trapping in a pair of periodically driven Rydberg atoms}
\author{S. Kumar Mallavarapu}
\affiliation{Department of Physics, Indian Institute of Science Education and Research, Dr. Homi Bhabha Road, Pune- 411008, Maharashtra, India}
\affiliation{Division of Physics, School of Advanced Sciences, Vellore Institute of Technology, Chennai, 600127, India}
\author{Ankita Niranjan}
\affiliation{Department of Physics, Indian Institute of Science Education and Research, Dr. Homi Bhabha Road, Pune- 411008, Maharashtra, India}
\author{Weibin Li}
\affiliation{School of Physics and Astronomy, University of Nottingham, Nottingham, NG7 2RD, United Kingdom}
\author{Sebastian W\"uster}
\affiliation{Department of Physics, Indian Institute of Science Education and Research, Bhopal, Madhya Pradesh, 462023,  India}
\author{Rejish Nath} 
\affiliation{Department of Physics, Indian Institute of Science Education and Research, Dr. Homi Bhabha Road, Pune- 411008, Maharashtra, India}

\begin{abstract}
We study the population trapping extensively in a periodically driven Rydberg pair. The periodic modulation of the atom-light detuning effectively suppresses the Rabi couplings and, together with Rydberg-Rydberg interactions, leads to the state-dependent population trapping. We identify a simple yet a general scheme to determine population trapping regions using driving induced resonances, the Floquet spectrum, and the inverse participation ratio.  Contrary to the single atom case, we show that the population trapping in the two-atom setup may not necessarily be associated with level crossings in the Floquet spectrum. Further, we discuss under what criteria population trapping can be related to dynamical stabilization, taking specific and experimentally relevant initial states, which include both product and the maximally entangled Bell states. The behavior of the entangled states is further characterized by the bipartite entanglement entropy. 
\end{abstract}



\maketitle

\section{Introduction}
Periodic driving emerged as a tool to coherently manipulate the states of quantum systems. Consequently, Floquet systems exhibit a wide variety of unique phenomena related to non-equilbirium dynamics and many-body physics \cite{she10, gol14, pon15, aba16, sil17, eck17, dal13, buk15}. One such phenomenon, the dynamical stabilization, has been a subject of study in both classical and quantum mechanical systems. Dynamical stabilization is the stabilization of an otherwise dynamically unstable  configuration of a system by periodically varying the system parameters in time. It has been first demonstrated using a classical pendulum, by Kapitza \cite{kap51}. By periodically moving the point of suspension with high frequency, it is possible to stabilize the pendulum in its inverted position. In the quantum world, a phenomenon closely analogous to the Kapitza pendulum is the population trapping in a two-level atom \cite{aga94, gar97, noe98}. The population can be trapped for a substantial time in an initial quantum state by periodically varying the atom-field detuning in time, even in cases where the state would otherwise evolve instantly into another state due to the Rabi coupling. Effectively, the periodic modulation may suppress the Rabi coupling depending on the modulation amplitude and frequency, leading to dynamical stabilization of the initial state. Dynamical stabilization has various applications, for instance, in ion-trapping \cite{pau90}, mass spectrometers, and particle synchrotrons \cite{cou52}.

Other quantum phenomena related to dynamical stabilization  are coherent destruction of tunneling in a double-well potential \cite{gri98, gro91,kie08}, the localization of a moving charged particle under the action of a time-periodic electric field \cite{dun86, rag96}, and the localization of a wavepacket in a periodic lattice due to periodic shaking of the lattice \cite{lig07,eck05,zen09, cre10} or modulating the inter-particle interactions \cite{gon09}.  In interacting quantum gases, a Kapitza or a dynamically stabilized state has different manifestations, for instance, stabilizing a Bose-Einstein condensate \cite{abd03} or a bright soliton \cite{sai03, sai07} against collapse, freezing spin mixing dynamics in spinor condensates \cite{zha10, hoa13, hua17}, inhibiting dissipation from a spin-half particle \cite{che15}, stabilizing a classically unstable phase ($\pi$-mode) in a bosonic Josephson junction \cite{bou10}, or giving rise to unconventional ordered phases that have no equilibrium counterparts \cite{ler19}. Additionally, dynamical stabilization has been used to control the superfluid-Mott insulator quantum phase transition of bosons in an optical lattice \cite{zen09}.

Currently, ultracold Rydberg atoms are emerging as a promising platform for probing quantum many-body phenomena and implementing quantum information protocols  \cite{bro20, saf10}. The Rydberg blockade, in which strong Rydberg-Rydberg interactions (RRIs) suppress simultaneous excitation of two Rydberg atoms within a finite volume \cite{luk01,urb09,gae09,hei07}, and the breaking of the blockade (anti-blockade) \cite{ate07,qia09,amt10,ssu20} are of central utility for these applications. For two atoms, it has been predicted that through modulation induced resonances, one can engineer the parameter space for both Rydberg-blockade and anti-blockade \cite{bas18, ank20, wuj20}. The latter is proposed to have applications in implementing robust quantum gates \cite{hua18, wu20, wuj20} and accelerating the formation of dissipative entangled steady states \cite{rui20}. To realize periodic driving in a Rydberg chain, either one can modulate the light field that couples the ground to the Rydberg state or applying additional radio-frequency fields. Those approaches give rise to sidebands either in the driving field or in the atomic levels \cite{aut55, gal08, noe98, mil16}.  Experiments with interacting Rydberg atoms in oscillating electric fields \cite{zhe15} have been employed to explore dipole-dipole interactions via F\"orster resonances \cite{tre14, zhe215, tau08, boh07}. Also, the dynamical stabilization of thermal Rydberg atoms against ionization, exposed to periodic kicks, has been a subject of intense study in the past, especially in classical-quantum correspondence \cite{yos00,rei97}. In the latter case, the RRIs were not relevant.  In a recent experiment, intensity-modulated off-resonance laser is used to vary the energy of an excited atomic state sinusoidally to generate interacting Rydberg polaritons \cite{cla19}.

In this paper, we study the population trapping comprehensively in a pair of periodically driven interacting two-level atoms, in which one of the energy levels is a Rydberg state. In particular, we consider the periodic modulation of the atom-field detuning. In general, the periodic modulation can enhance or suppress the population dynamics in the system, and the latter implies population trapping. In our setup, the dynamical stabilization emerges as a particular case of population trapping. The stability  of the initial state against "time evolution" decides whether the population trapping can be accounted as the phenomenon of dynamical stabilization. Suppose the initial state evolves in time in the absence of periodic modulation. In that case, we have the dynamical stabilization phenomenon similar to that of the Kapitza pendulum. Whenever this scenario occurs, we explicitly state it as dynamical stabilization; otherwise, we use the term population trapping. 

The two-atom setup we consider is one of the most common scenarios in Rydberg atom experiments \cite{gae09, wil10,ise10,rya10, beg13, rav14, lab14,rav15,jau15,del17,zen17,pic18,lev18} and can be easily realizable using optical tweezers or microscopic optical traps \cite{beg13}. The same setup also constitutes the basic building block for quantum simulations and quantum information protocols \cite{saf10}. We show that the presence of RRIs leads to (initial) state-dependent population trapping in the modulated two-atom setup. In particular, we look at how a specific set of experimentally relevant initial states, including both product and maximally entangled Bell states, can be dynamically stabilized or freeze for significantly long periods. The product states we consider are those in which both atoms occupy either ground or Rydberg states. In a Rydberg setup, the Bell states have been demonstrated experimentally using various techniques \cite{wil10,ise10,zhan10,mal15,jau15,omr19}. We identify a simple scheme for locating population trapping regions for any initial state, relying on driving induced resonances and the Floquet spectrum. We also introduce inverse participation ratio (IPR), calculated from the overlap of the initial state with the Floquet eigenstates, as an indicator of population trapping. Contrary to the previous conception from the single atom case, the population trapping or the dynamical stabilization in the two-atom setup is not necessarily related to the level crossings in the Floquet spectrum.
 
The paper is structured as follows. In Sec. \ref{set}, we discuss the physical setup, the Hamiltonians including an effective time-independent one in the high-frequency limit, and techniques which we employ to study the emergence of Kapitza or dynamically stabilized states. The population trapping including the dynamical stabilization in a single two-level atom and the scheme for identifying dynamical stabilization are discussed in Sec.~\ref{n1}. In Sec.~\ref{n2}, we extend the scheme to the two atom setup, and in particular, discuss the population trapping in both product and entangled states, including the driving induced resonances, and the Floquet spectrum. Finally, we summarize in Sec.~\ref{sum}.

\section{Setup, Model, and Techniques}
\label{set}
We consider a chain of two two-level atoms, in which the electronic ground state $|g\rangle$ is coupled to a Rydberg state $|e\rangle$ via a light field, the frequency of which is varied periodically in time $t$. The system is described in the frozen gas limit, after the rotating wave and dipole approximations, by the time-dependent Hamiltonian ($\hbar=1$):
\begin{equation}
\hat H=-\Delta(t)\sum_{i=1}^2\hat\sigma_{ee}^{i}+\frac{\Omega}{2}\sum_{i=1}^2\hat\sigma_x^{i}+V_0\hat\sigma_{ee}^{1}\hat\sigma_{ee}^{2},
\label{ham}
\end{equation}
where $\hat\sigma_{ab}=|a\rangle\langle b|$ with $a, b\in \{e, g\}$ includes both transition and projection operators, $\hat\sigma_x=\hat\sigma_{eg}+\hat\sigma_{ge}$, $\Omega$ is the Rabi frequency, $\Delta(t)=\Delta_0+\delta\sin\omega t$ is the time-dependent detuning with modulation amplitude $\delta>0$ and the modulation frequency $\omega$. The Rydberg excited atoms interact via strong van der Waals interactions, $V_0=C_6/r^6$, where $C_6$ is the interaction coefficient, and $r$ is the separation between two Rydberg excitations \cite{beg13}. The exact dynamics of the system is obtained by numerically solving the Schr\"odinger equation: $i\partial \psi(t)/\partial t=\hat H(t)\psi(t)$. To gain an insight, especially at high modulation frequency ($\omega\gg\Omega$), we move to a rotating frame: $|\psi'\rangle=\hat U(t)|\psi\rangle$ where $\hat U(t)=\exp[if(t)\sum_j\hat\sigma_{ee}^{j}+itV_0\hat\sigma_{ee}^{1}\hat\sigma_{ee}^{2}]$ with $f(t)=(\delta/\omega)\cos\omega t-\Delta_0 t$.  The new Hamiltonian, $\hat H'(t)=\hat U\hat H\hat U^{\dagger}-i\hbar\hat U\dot{\hat U}^{\dagger}$, after using the Jacobi-Anger expansion $\exp({\pm iz\cos\omega t})=\sum_{m=-\infty}^{\infty}J_m(z)\exp(\pm im[\omega t+\pi/2])$, is \cite{bas18}
\begin{eqnarray}
\hat H'&=&\frac{\Omega}{2}\sum_{j=1}^2\sum_{m=-\infty}^{\infty}i^mJ_m(\alpha)g_m(t)e^{iV_0\sum_{k\neq j}\hat\sigma_{ee}^{k}t}\hat\sigma_{eg}^{j}+{\rm H.c.} 
\label{hamro}
\end{eqnarray}
where $J_m(\alpha)$ is the $m$th order Bessel function with $\alpha=\delta/\omega$ and $g_m(t)=\exp[i(m\omega-\Delta_0)t]$. Comparing Eq.~(\ref{ham}) with Eq.~(\ref{hamro}), we can see that the periodic detuning has effectively modified the Rabi coupling, thereby affecting the excitation dynamics. Further, using $e^{\pm iV_{0}\sum_{k\neq j}\hat\sigma_{ee}^{k}t}=\prod_{k\neq j}  \left[\hat\sigma_{ee}^k (e^{\pm itV_{0}}-1)+\mathcal I\right]$, where $\mathcal I$ is the identity operator, we rewrite the Hamiltonian in Eq.~(\ref{hamro}) as 
\begin{equation}
 \hat H'=\frac{\Omega}{2}\sum_{m=-\infty}^{\infty}i^m J_m(\alpha)g_m(t)\left(\sum_{j=1}^2\hat\sigma_{eg}^j+\hat X\left(e^{iV_0t}-1\right)\right)+ {\rm H.c.},
 \label{hro2}
 \end{equation}
where the operator $\hat X=\hat\sigma_{eg}^1\hat\sigma_{ee}^2+\hat\sigma_{eg}^2\hat\sigma_{ee}^1$ describes the correlated Rabi coupling \cite{bas18, sri19}. The correlated Rabi process is analogous to the density assisted inter-band tunneling or density-dependent hopping for atoms in optical lattices \cite{cho16,dut15}.


{\em Floquet Theory}.--- According to the Floquet theorem, the time evolution operator associated with a time-periodic Hamiltonian $\hat H(t)$ is $\hat U(t)=P(t)e^{-i\hat H_Ft}$, where the Floquet Hamiltonian $\hat H_F$ is defined through the evolution operator over a full period $T=2\pi/\omega$, i.e., $\hat U(T)=e^{-i\hat H_FT}$  \cite{shi65, sam73, gri98, buk15, mar20}. The unitary operator $\hat P(t)=\hat P(t+T)$ has the same periodicity as that of the Hamiltonian, and it becomes an identity operator at the instants $t_n=nT$ where $n=0, 1, 2, ...$. Further, we can write, $\hat U(T)=e^{-i\hat H_FT}=\sum_ke^{-i\theta_k}|\phi_k(0)\rangle\langle\phi_k(0)|$, where the Floquet modes $\{|\phi_{k}(0)\rangle\}$ are the eigenstates of the Hamiltonian $\hat H_F$, and they form a complete set of square-integrable states. The Floquet mode $|\phi_{k}(t) \rangle=\exp(\mathrm{i} \epsilon_{k} t)\hat U(t)|\phi_{k}(0)\rangle$ has the same periodicity in time as that of the Hamiltonian $\hat H(t)$, and the quasi-energy $\epsilon_k=\theta_k/T$ is defined up to a multiple of $\omega$.  Then, a general state of the system can be written as
\begin{equation}
|\psi(t) \rangle =  \sum_{k} c_{k} \mathrm{exp}(-\mathrm{i} \epsilon_{k} t)|\phi_{k}(t) \rangle,
\label{tf}
\end{equation}
where the time-independent co-efficient $c_k$ gives the probability amplitude for finding the system in the Floquet mode $|\phi_{k}(t) \rangle$ and is determined from the initial state $|\psi(0) \rangle$. It is worth mentioning that the population in the Floquet modes remains preserved even if the actual state of the system or the Hamiltonian is changing over time. In that spirit, if the initial state coincides with one of the Floquet modes, the population trapping takes place. The quasi-energies $\epsilon_k$ and the modes $\{|\phi_{k}(0)\rangle\}$ are calculated numerically by obtaining the eigenvalues, $\lambda_k=\exp(-i\epsilon_kT)$ of the one-period operator $\hat U(T)$ \cite{cre02, cre03}. To obtain $\hat U(T)$, we evolve each of the basis states using the original Hamiltonian in Eq.~(\ref{ham}).

Further, to characterize the behavior of Rydberg excitation dynamics we define the inverse participation ratio (IPR),
\begin{equation}
\Pi_N^{|I\rangle}=\frac{1}{\sum_k p_k^2}-1,
\end{equation}
where $p_k=|\langle\phi_k(0)|I\rangle|^2$, is the projection of the initial state $|I\rangle$ on the Floquet mode $|\phi_k(0)\rangle$ for $N$ atoms. If the initial state coincides with one of the Floquet modes, IPR vanishes. Since, the population in Floquet mode doesn't vary in time, $\Pi_N^{|I\rangle}=0$ may indicate the population trapping or dynamical stabilization of the state $|I\rangle$. In the same spirit, a smaller value of $\Pi_N^{|I\rangle}$ indicates a slower  transition rate from the state $|I\rangle$ to other states.

\section{A two-level atom ($N=1$)}
\label{n1}
\begin{figure*}
\centering
\includegraphics[width= 2.\columnwidth]{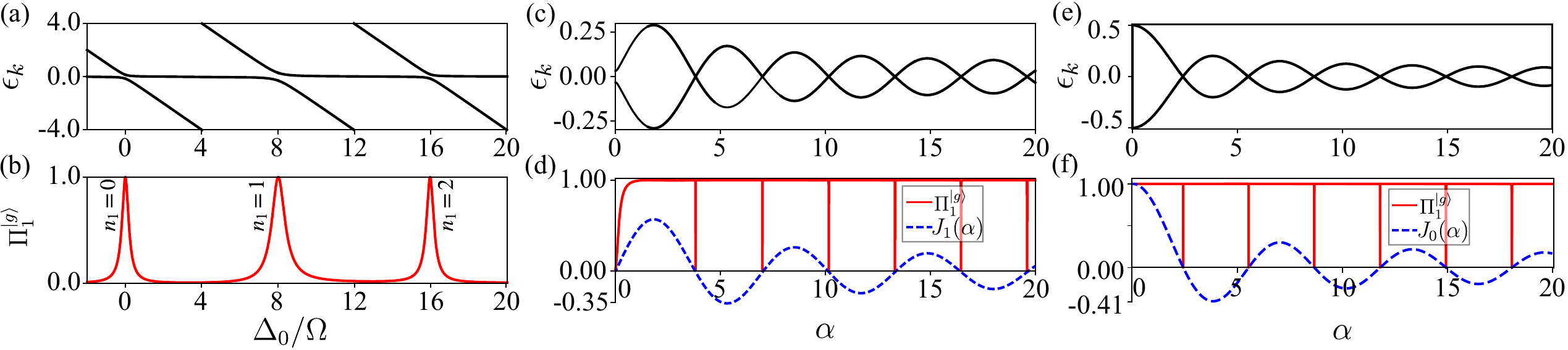}
\caption{\small{(Color online) Floquet mode properties of a driven single two-level atom with $\omega=8\Omega$. (a) The quasi-energies $\epsilon_k$  and (b) IPR ($\Pi_1^{|g\rangle}$) as a function of $\Delta_0$ for $\delta=15\Omega$. (c) The quasi-energies $\epsilon_k$  and (d) IPR ($\Pi_1^{|g\rangle}$) as a function of $\alpha=\delta/\omega$ for the resonance $\Delta_0=\omega$ ($n_1=1$). In (d), we also show the Bessel function, $J_1(\alpha)$. Its zeros coincide with $\Pi_1^{|g\rangle}=0$ indicating the population trapping. The parameter $\alpha$ is varied by changing $\delta$. (e) and (f) show the results for the case of primary resonance, $\Delta_0=0$ ($n_1=0$). The crossings of $\epsilon_k$ in (e) and the zeros of $\Pi_1^{|g\rangle}$ in (f) coincides with the zeros of $J_0(\alpha)$. The plots (e) and (f) are the special case of population trapping corresponding to the dynamical stabilization. The parameter $\alpha$ is varied by changing $\delta$ and keeping $\omega$ constant.}}
\label{fig:1} 
\end{figure*}
In the following, we briefly review the population trapping in a periodically driven single two-level atom. In particular, we discuss the criteria under which the population trapping can be identified as dynamical stabilization. For $N=1$, the Hamiltonian in Eq.~(\ref{hamro}) takes the simplest form \cite{aga94, gar97, noe98}, 
\begin{equation}
 \hat H'=\frac{\Omega}{2}\sum_{m=-\infty}^{\infty}i^m g_m(t)J_m(\alpha)\hat\sigma_{eg}+ {\rm H.c.}.
 \label{ham1}
 \end{equation}
In the high-frequency limit ($\omega\gg\Omega$), the terms satisfying the resonance condition, $n_1\omega=\Delta_0$, where $n_1=0, 1, 2, ...$ becomes the only relevant term in the summation of Eq.~(\ref{ham1}). Neglecting non-resonant terms is equivalent to a second rotating wave approximation. Once the resonance condition is satisfied, the population dynamics exhibits coherent Rabi oscillations between $|g\rangle$ and $|e\rangle$. In Figs.~\ref{fig:1}(a) and \ref{fig:1}(b), we show the Floquet spectrum and IPR ($\Pi_1^{|g\rangle}$) as a function of $\Delta_0$. The resonances can be identified as either avoided crossings in the Floquet spectrum or peaks in the IPR ($\Pi_1^{|g\rangle}$). At those peaks ($\Pi_1^{|g\rangle}=1$), the Floquet modes become an equal superposition of $|g\rangle$ and $|e\rangle$. Far away from the avoided crossings (resonances), i.e., for $\Delta_0\neq n_1\omega$ and $\Delta_0\gg\Omega$, the periodic driving is ineffective. In that case, the Floquet modes approximately become the eigenstates of the undriven Hamiltonian, $\hat H(t=0)$, which are either $|g\rangle$ or $|e\rangle$ with a weak mixing between them. Due to this, $\Pi_1^{|g\rangle}$ decays to almost zero between the resonances. 

\begin{figure}
\centering
\includegraphics[width= 1.\columnwidth]{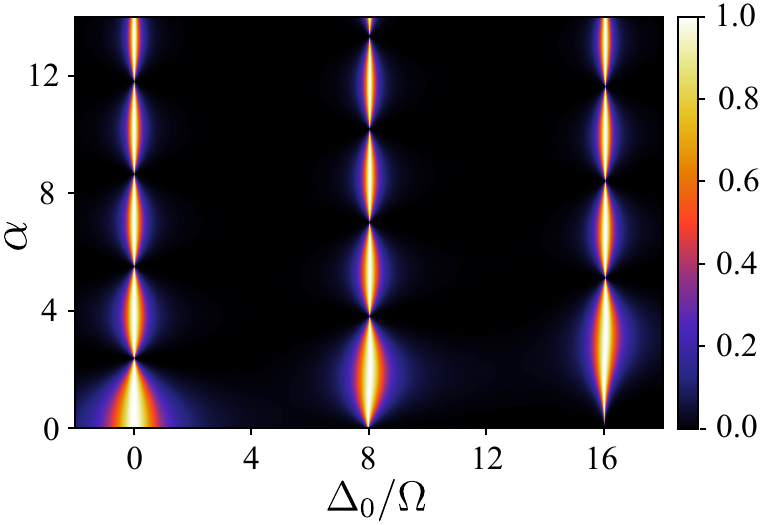}
\caption{\small{(Color online) The IPR ($\Pi_1^{|g\rangle}$) as a function of $\alpha$ and $\Delta_0$ for $\omega=8\Omega$. The pearl-stripes are along the $\alpha$ axis at the resonances $n\omega=\Delta_0$. The local minima ($\Pi_1^{|g\rangle}=0$) along the first stripe are the points of DS for which $J_0(\alpha)$=0. The parameter $\alpha$ is varied by changing $\delta$ and keeping $\omega$ constant.}}
\label{fig:2} 
\end{figure}

At the resonance $n_1\omega=\Delta_0$, the effective Rabi coupling between the states $|g\rangle$ and $|e\rangle$ is proportional to $J_{n_1}(\alpha)$. Therefore, at the Bessel zeros [$J_{n_1}(\alpha)=0$], the dynamics freezes and leads to population trapping. This can be further verified by looking at the quasi-energies $\epsilon_k$ as a function of $\alpha$ keeping the resonance condition satisfied. The quasi-energies or the energy gap between them oscillate as a function of $\alpha$, and crossings occur at the zeros of the Bessel function \cite{cre03}. Fig.~\ref{fig:1}(c) and \ref{fig:1}(d) show the results for the case of $\Delta_0=\omega$, and the crossings occur at the zeros of $J_{1}(\alpha)$. At those crossings, the degenerate Floquet modes become purely $|g\rangle$ and $|e\rangle$, which results in a vanishing $\Pi_1^{|g\rangle}$ or $\Pi_1^{|e\rangle}$ as seen in Fig.~\ref{fig:1}(d). Since the Floquet modes do not evolve in time, the population in states $|g\rangle$ or $|e\rangle$ freezes.  Note that at the crossings, an arbitrary superposition of $|g\rangle$ and $|e\rangle$ is also a Floquet mode making the population trapping independent of the initial state. As we show below, the latter breaks down in the presence of RRI, leading to a state-dependent population trapping. In short, a vanishing IPR at the driving induced resonance indicates the freezing of the initial state or population trapping. 

Note that only if the initial state is dynamically unstable in the absence of periodic modulation, then only the corresponding population trapping can be called the dynamical stabilization. It is easy to see that the dynamical stabilization occurs only when $n_1=0$. If $n_1$ is a non-zero integer, in the high-frequency limit, the resonance condition demands a large value of $\Delta_0$. For such large values of $\Delta_0$, there is hardly any dynamics in the states $|g\rangle$ and $|e\rangle$ in the absence of periodic driving. Therefore, population trappings for $n_1>0$ cannot be interpreted as dynamical stabilization. In other words, the population trapping at the primary resonance ($n_1=0$), i.e., when $J_0(\alpha)=0$ for $\Delta_0=0$, provides us the phenomenon of dynamical stabilization. The results for the latter case with an initial state $|I\rangle=|g\rangle$ are shown in Figs.~\ref{fig:1}(e) (quasi-energies) and \ref{fig:1}(f) (IPR). Note that the leading terms in the excited state population due to $m\neq n_1$ terms in Eq.~(\ref{ham1}) are proportional to $(\Omega/\omega)^2$ in the high-frequency limit, which can be ignored \cite{gar97}. More extensive results of the IPR ($\Pi_1^{|g\rangle}$) for the initial state $|g\rangle$, are given in Fig.~\ref{fig:2}. In the $\alpha-\Delta_0$ plane, $\Pi_1^{|g\rangle}$ exhibits pearl-chains along $\alpha$ axis at the resonances $n_1\omega=\Delta_0$. The local minima along the chains provide the values of $\alpha$ at which population trapping takes place [or $J_n(\alpha)=0$], and those along $\alpha$ at $\Delta_0=0$ are the points of dynamical stabilization.  Between the stripes (along $\Delta_0$ axis), $\Pi_1^{|g\rangle}$ vanishes due to the far off-resonant driving of the atom, as discussed above. Note that the effect of a finite $\omega$ is apparent only for sufficiently small $\omega$ for which the crossings in Floquet energies start to deviate slightly from the Bessel zeros.

In short, by varying the amplitude of periodic modulation, the avoided crossings (resonances) [see Fig.~\ref{fig:1}(a)] in the quasi-energy spectrum become actual level crossings [see Fig.~\ref{fig:1}(c)]. At the crossings, the population dynamics freezes, and also the IPR vanishes. We term this, at resonance, as Population trapping. Population trapping at the primary resonance is identified as the dynamical stabilization. Thus, we have a scheme to identify population trapping (including dynamical stabilization) of any initial state in two steps. First, identify resonances in which the initial state is involved, and second, vary the amplitude of modulation, keeping the resonance condition satisfied.

\hspace{0.1 cm}
\section{Two-atom chain ($N=2$)}
\label{n2}
This section extends the above analysis from a single two-level atom to a pair of Rydberg atoms and discusses how RRIs affect the population trapping. In particular, we are interested in the conditions under which the states $|gg\rangle$, $|ee\rangle$, $|+\rangle=(|eg\rangle+|ge\rangle)/\sqrt{2}$, and $|B\rangle=(|gg\rangle+|ee\rangle)/\sqrt{2}$ are dynamically stabilized. The first two states are product states, and the last two are the maximally entangled Bell states. If we restrict the dynamics to the symmetric states, we can truncate the basis to $\{|gg\rangle, |+\rangle, |ee\rangle\}$. On this basis, the off-diagonal matrix elements of $\hat H'$ in Eq.~(\ref{hro2}) provide the time-dependent coupling strengths for $|gg\rangle\leftrightarrow|+\rangle$ and $|+\rangle\leftrightarrow|ee\rangle$ transitions, and they are respectively,
 \begin{eqnarray}
    \label{c1}
    \Omega_1(t) &\propto&  \frac{\Omega}{\sqrt{2}} \sum_{m = -\infty}^{\infty} J_m(\alpha) e^{i(m\omega-\Delta_0)t +im\pi/2} \\
    \Omega_2(t) &\propto&  \frac{\Omega}{\sqrt{2}}  \sum_{m = -\infty}^{\infty} J_m(\alpha) e^{i(m\omega-\Delta_0+V_0)t +im\pi/2},
    \label{c2}
\end{eqnarray}
and in general, $\Omega_1\neq \Omega_2$. As a first step towards analyzing the population trapping, we discuss the resonances in the two-atom driven setup.
\subsection{Resonances}
\begin{figure}
\centering
\includegraphics[width= 1.\columnwidth]{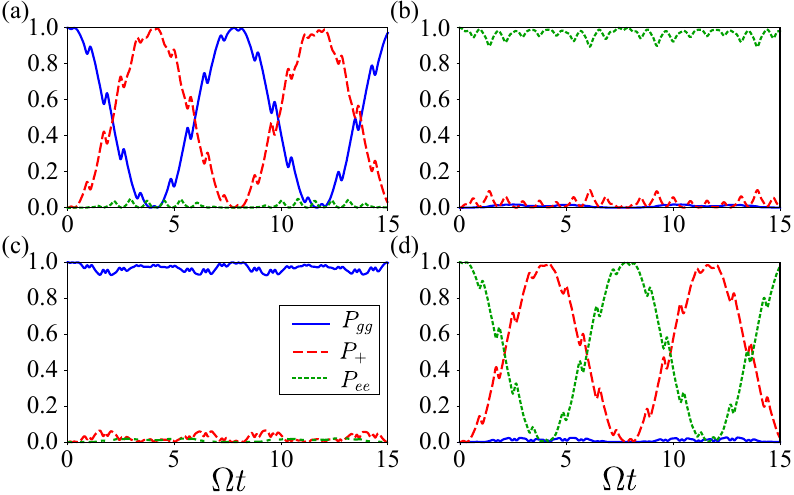}
\caption{\small{(Color online) Population dynamics for the resonance type R1 ($n_1\omega_0=\Delta_0$) for the initial states (a) $|I\rangle=|gg\rangle$ and (b) $|I\rangle=|ee\rangle$. The same, but with the resonance type R2 ($n_2\omega=\Delta_0-V_0$) for the initial state (c) $|I\rangle=|gg\rangle$ and (d) $|I\rangle=|ee\rangle$ with $\Delta_0=2\Omega$. In (a) we see the Rabi oscillations between $|gg\rangle$ and $|+\rangle$ states, whereas in (b) we observe no dynamics. Similarly, (c) shows the absence of dynamics, and the Rabi oscillations between $|+\rangle$ and $|ee\rangle$ states is shown in (d). We took $V_0=10 \Omega$, $\delta=15\Omega$, and $\omega=8 \Omega$ for all plots. The value of $\Delta_0$ is taken such that $n_1=1$ for (a) and (b), and for (c) and (d) we have $n_2=-1$.}}
\label{fig:3} 
\end{figure}

 At high $\omega$, the most relevant terms in Eqs.~(\ref{c1}) and (\ref{c2}) give the resonance criteria  $n_1\omega=\Delta_0$ (R1) and $n_2\omega=\Delta_0-V_0$ (R2), which are associated with the transitions $|gg\rangle\leftrightarrow|+\rangle$ and $|+\rangle\leftrightarrow|ee\rangle$, respectively. For sufficiently large values of  $|V_0-n\omega|$ with $n=0, \pm 1, \pm 2, ... $, the resonances of the types R1 and R2 can be well separated along the $\Delta_0$ axis. If $V_0=n\omega$, the criteria for R1 and R2 are satisfied simultaneously with $n_1=n_2+n$. Assuming R1 and R2 resonances do not overlap, and only if R1 is fulfilled, the effective (time-averaged) Rabi couplings become $\Omega_1\approx\Omega J_{n_1}(\alpha)/\sqrt{2}$ and $\Omega_2\approx 0$, for $|gg\rangle\leftrightarrow|+\rangle$ and $|+\rangle\leftrightarrow|ee\rangle$ transitions, respectively. Therefore, for the initial state $|I\rangle=|gg\rangle$, the system exhibits Rabi oscillations between $|gg\rangle$ and $|+\rangle$ states [see Fig.~\ref{fig:3}(a) for $n_1=1$], which corresponds to the dynamics under the Rydberg blockade. In contrast, if $|I\rangle=|ee\rangle$, the dynamics freezes, as shown in Fig.~\ref{fig:3}(b). The latter is expected since the state $|ee\rangle$ is far off-resonant from $|+\rangle$ due to large $V_0$, and hence, the periodic driving is nonrelevant. If the condition for R2 is satisfied, we have $\Omega_1\approx 0$ and $\Omega_2\approx \Omega J_{n_2}(\alpha)/\sqrt{2}$ which leads to the Rabi oscillations between $|ee\rangle$ and $|+\rangle$ states and hardly any dynamics if the initial state is $|gg\rangle$, as shown in  Figs.~\ref{fig:3}(c) and \ref{fig:3}(d) for $n_2=-1$, respectively. Apart from the resonances R1 and R2, there exists a third one $n_3\omega=2\Delta_0-V_0$ (R3), which is not directly visible from Eqs.~(\ref{c1}) and (\ref{c2}), but can be revealed using adiabatic impulse approximation \cite{ank20}. R3 leads to resonant transitions between $|gg\rangle$ and $|ee\rangle$. 

\begin{figure}
\vspace{0.cm}
\centering
\includegraphics[width= 1.\columnwidth]{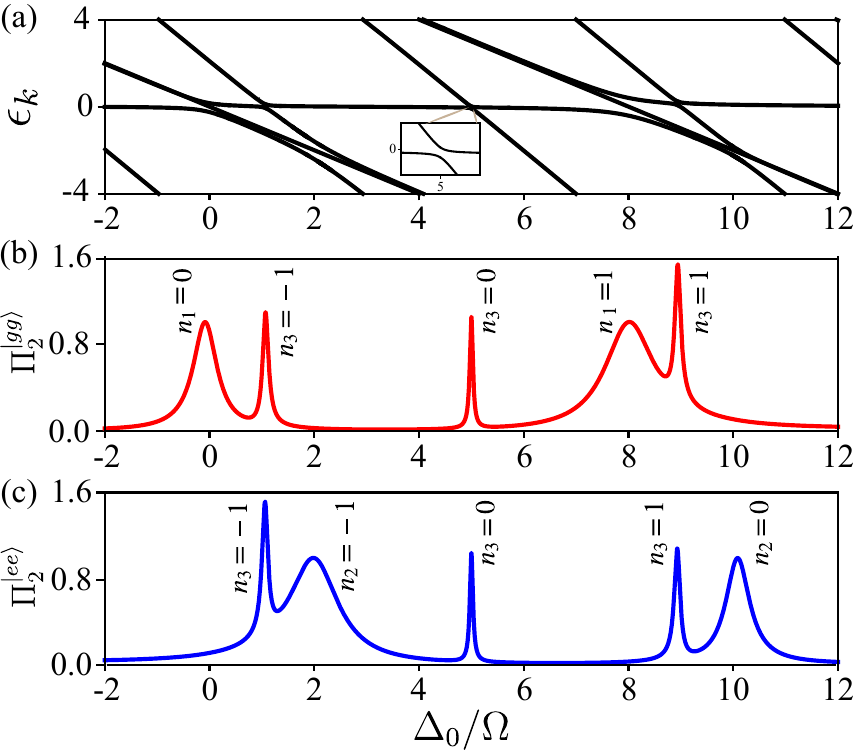}
\caption{\small{(Color online)(a) The quasi-energy spectrum for $N=2$ as a function of $\Delta_0$ for $V_0=10 \Omega$, $\delta=15\Omega$, and $\omega=8 \Omega$. (b) and (c) show $\Pi_2^{|gg\rangle}$ and $\Pi_2^{|ee\rangle}$, respectively. The peaks in $\Pi_2$ and the avoided crossings in $\epsilon_k$ indicate the three different resonant transitions: (R1) $n_1\omega=\Delta_0$, (R2) $n_2\omega=\Delta_0-V_0$, and (R3) $n_3\omega=2\Delta_0-V_0$ labelled by $n_1$, $n_2$, and $n_3$, respectively.}}
\label{fig:4} 
\end{figure}

In Figs.~\ref{fig:4}(b) and \ref{fig:4}(c), we show the IPR ($\Pi_2^{|I\rangle}$) as a function of $\Delta_0$ for the initial states $|gg\rangle$ and $|ee\rangle$, respectively. The value of other parameters is the same as in Fig.~\ref{fig:3}. The peaks in Fig.~\ref{fig:4}(b) correspond to the resonances R1 and R3, labeled by $n_1$ and $n_3$, respectively. Similarly, the peaks in Fig.~\ref{fig:4}(c) correspond to the resonances R2 and R3, labeled by $n_2$ and $n_3$, respectively. As expected, the R3 resonances (marked by $n_3$) are very narrow since $|gg\rangle$ and $|ee\rangle$ are not directly coupled. Between the resonant peaks, $\Pi_2^{|I\rangle}$ vanishes due to the off-resonant driving as discussed above. These resonances cause the avoided crossings in the quasi-energies shown in Fig.~\ref{fig:4}(a). To calculate $\epsilon_k$ in Fig.~\ref{fig:4}(a), we used the basis $\{|gg\rangle, |eg\rangle, |ge\rangle, |ee\rangle\}$ and therefore we have four levels in the quasi-energy spectrum.

\subsection{Dynamical stabilization of product states: $|gg\rangle$ and $|ee\rangle$}
\label{ps}
\begin{figure}
\centering
\includegraphics[width= .9\columnwidth]{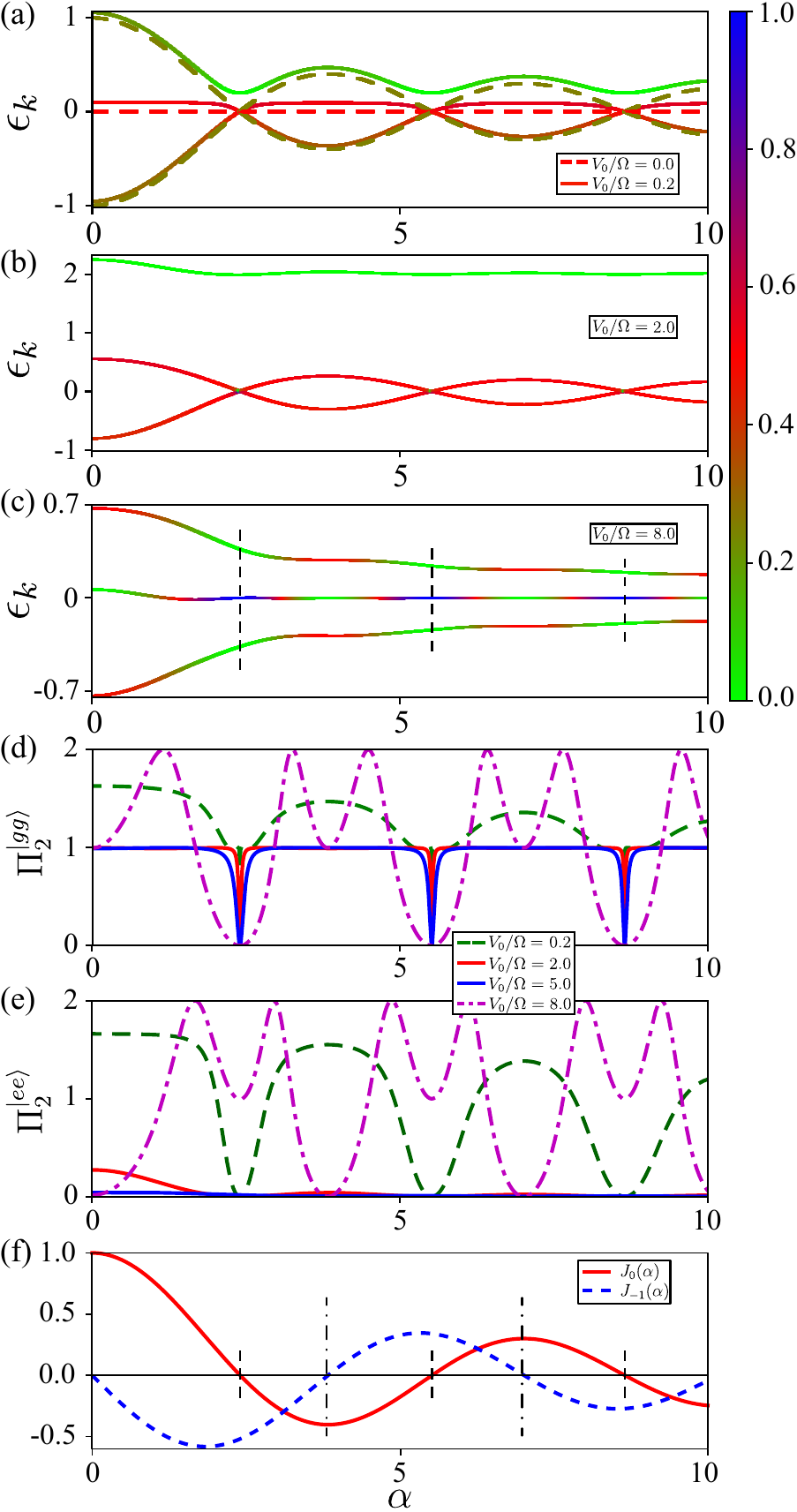}
\caption{\small{(Color online) The quasi-energy spectrum $\epsilon_k$ and IPR ($\Pi_2^{|gg\rangle}$, $\Pi_2^{|ee\rangle}$) for $N=2$, $\Delta_0=0$, and $\omega=8\Omega$, as a function of $\alpha$ for different $V_0$. (a) shows $\epsilon_n$ for $V_0=0 \Omega$ (dashed lines), and  $V_0=0.2 \Omega$ (solid lines), and (b) and (c) show the same for $V_0=2 \Omega$ and $V_0=8 \Omega$, respectively. Since $\Delta_0=0$, in (a) and (b), the level crossings take place at the zeros of $J_0(\alpha)$. In (a)-(c) the color bar indicates the probability of the finding the state $|gg\rangle$ in each of the Floquet modes. The dashed vertical lines in (c) mark $J_0(\alpha)=0$, and at those points the central Floquet mode consists purely of $|gg\rangle$ state, which indicates dynamical stabilization. (d) and (e) show the IPR $\Pi_2^{|gg\rangle}$ and $\Pi_2^{|ee\rangle}$, respectively. In (f), we show the Bessel functions $J_0(\alpha)$ (solid line) and $J_{-1}(\alpha)$ (dashed line). The parameter $\alpha$ is varied by changing $\delta$ and keeping $\omega$ constant.}}
\label{fig:5} 
\end{figure}

{\em R1}.--- First we discuss the dynamical stabilization of the product states $|gg\rangle$ and $|ee\rangle$. To identify the regions of dynamical stabilization we choose the primary resonance in each of R1, R2 and R3, i.e., $n_{j\in 1, 2, 3}=0$ and vary the amplitude of modulation. Equivalently, one can vary $\alpha$ by keeping $\omega$ constant. First, we consider the resonance R1 with $n_1=0$ ($\Delta_0=0$). For the non-interacting case ($V_0=0$), as discussed for the single atom case in Sec.~\ref{n1}, the dynamical stabilization occurs at the zeroes of the $J_{0}(\alpha)$. As expected, when $J_{0}(\alpha)=0$, all three quasi-energies cross [dashed lines in Fig.~\ref{fig:5}(a)]. Since we have eliminated the asymmetric state $|-\rangle=(|eg\rangle-|ge\rangle)/\sqrt{2}$ from the dynamics, there are only three relevant quasi-energy eigenvalues. The color bar in Fig.~\ref{fig:5} quantifies the probability density of $|gg\rangle$ in each of the Floquet states. A finite $V_0$ partially lifts the degeneracy of $\epsilon_k$ at the crossings [see solid lines in Fig.~\ref{fig:5}(a)]. For small RRIs ($V_0\ll\Omega$), the resonance R2 is not well isolated from R1 and all three states ($|gg\rangle$, $|+\rangle$, $|ee\rangle$) participate in the dynamics for any initial state. Therefore, we need to address the dynamical stabilization of both $|gg\rangle$ and $|ee\rangle$ when RRI is small.

Satisfying R1 and for $V_0/\omega\ll 1$, in the high-frequency limit ($\omega\gg\Omega$), we can obtain an effective time-independent Hamiltonian from Eq.~(\ref{hro2}) as, $H_{{\rm eff}}=1/T\int_0^Tdt \ \hat H'(t)$ where $T=2\pi/\omega$ \cite{gon09, buk15, ale14} (see Appendix \ref{a1}). Then, expanding $\hat H_{{\rm eff}}$ in 
powers of $V_0/\omega$ we have,
\begin{eqnarray}
\hat H_{{\rm eff}}^{(V_0\ll\omega)}\simeq\frac{ i^{n_1}J_{n_1}(\alpha)\Omega}{2}\left(\sum_{j=1}^2\hat\sigma_{eg}^j+i\pi\frac{V_0}{\omega}\hat X\right)+\nonumber \\
\frac{\Omega}{2}\sum_{m\neq n_1}\frac{i^mJ_m(\alpha)}{(m-n_1)}\frac{V_0}{\omega}\hat X+\mathcal {O}\left (V_0^2/\omega^2\right)+{\rm H.c.}.
\label{heff3}
\end{eqnarray}
Equation (\ref{heff3}) implies that in the infinite-frequency limit ($V_0/\omega\to 0$), the population trapping occurs at the zeros of the Bessel function $J_{n_1}(\alpha)$ irrespective of the initial state. At the primary resonance ($n_1=0$), we have the dynamical stabilization. For non-zero, but small values of $V_0/\omega$, the dominant interaction dependence comes from the second and third terms in Eq.~(\ref{heff3}), which are linear in $V_0/\omega$. For $n_1=0$, the third term in Eq.~(\ref{heff3}) vanishes, which means that the DS occurs at $J_{0}(\alpha)=0$. To verify this, we analyze IPRs $\Pi_2^{|gg\rangle}$ and $\Pi_2^{|ee\rangle}$ as a function of $\alpha$, shown respectively in Figs.~\ref{fig:5}(d) and \ref{fig:5}(e) for $V_0=0.2 \Omega$ and $\Delta_0=0$ (green dashed lines). As expected, they both vanish when $J_0(\alpha)=0$, indicating the dynamical stabilization of both $|gg\rangle$ and $|ee\rangle$.

 When $n_1\neq 0$, and for $\alpha$ such that $J_{n_1}(\alpha)=0$, the third term in Eq.~(\ref{heff3}) also becomes vanishingly small and can be safely ignored. That means, for small values of $V_0/\omega$ with R1 being satisfied, the population trapping always occurs at the zeros of the Bessel function $J_{n_1}(\alpha)$. The corrections from the terms involving $\hat X$ in Eq.~(\ref{heff3}) may introduce a tiny shift in the value of $\alpha$ at which the DS occurs, especially for the case, $|I\rangle=|ee\rangle$. It can also be seen from Fig.~\ref{fig:5}(a) that the value of $\alpha$ for which the crossings in the Floquet spectrum occur hardly affected by small values of $V_0$. 

Coming back to the case of dynamical stabilization for $n_1=0$ and as $V_0$ increases (excluding $V_0=n\omega$ where $n$ is a non-zero positive integer), one quasi-energy level [topmost level in Figs.~\ref{fig:5}(a) and \ref{fig:5}(b)] moves away from the other two, and eventually becomes purely $|ee\rangle$ in the blockade regime ($V_0\geq \Omega$), for any value of $\alpha$ [see Fig.~\ref{fig:5}(b)]. At that stage, the two lowest Floquet modes shown in Fig.~\ref{fig:5}(b) become superposition of $|gg\rangle$ and $|+\rangle$ states, except at the level crossings. At the crossings, which occur for $J_0(\alpha)=0$, the two Floquet modes become purely $|gg\rangle$ and $|+\rangle$ states, and $|gg\rangle$ is dynamically stabilized. The latter is further confirmed by $\Pi_2^{|gg\rangle}$ [see Fig.~\ref{fig:5}(d).], which vanishes at the crossings. $\Pi_2^{|gg\rangle}=1$ implies Rydberg blockade for which we have an effective two-level system consisting of $|gg\rangle$ and $|+\rangle$ states. In the blockade regime, the state $|ee\rangle$ is dynamically stable even in the absence of periodic driving, which makes $\Pi_2^{|ee\rangle}\sim 0$ independent of $\alpha$ [see Fig.~\ref{fig:5}(e) for $V_0=5\Omega$]. 

When $V_0=n\omega$, where $n$ is a non-zero positive integer, both R1 and R2 are satisfied simultaneously. In that case, the Bessel functions $J_{n_1}(\alpha)$ and  $J_{n_2=n_1-n}(\alpha)$ [see Eqs.~(\ref{c1}) and (\ref{c2})] determine the couplings for the transitions $|gg\rangle\leftrightarrow|+\rangle$ and $|+\rangle\leftrightarrow|ee\rangle$, respectively. In Figs.~\ref{fig:5}(c)-\ref{fig:5}(e), we show the results for $\Delta_0=0$ and $V_0=\omega=8\Omega$, therefore $n_1=0$ and $n_2=n_1-n=-1$. Thus, the dynamical stabilization of $|gg\rangle$ occurs at the zeros of $J_{0}(\alpha)$, and the population trapping in $|ee\rangle$ takes place when $J_{-1}(\alpha)=0$. When both R1 and R2 are satisfied simultaneously, both $\epsilon_k$ and $\Pi_2^{|gg\rangle}$ exhibit qualitatively different features compared to the case when only either R1 or R2 (see below) is satisfied. The first thing to notice is that $\epsilon_k$ does not show any level crossings as a function of $\alpha$ [see Fig.~\ref{fig:5}(c)]. Despite that, we observe dynamical stabilization of $|gg\rangle$ at $J_{0}(\alpha)=0$ [marked by dashed vertical lines in Fig.~\ref{fig:5}(c)]. Because at those values of $\alpha$, one of the Floquet modes [middle one in Fig.~\ref{fig:5}(c)] becomes purely $|gg\rangle$. It is in stark contrast to the case of a single two-level atom for which the dynamical stabilization is always accompanied by a level crossing in the quasi-particle spectrum. Additionally, both $\Pi_2^{|gg\rangle}$ and $\Pi_2^{|ee\rangle}$ exhibit primary and secondary minima as a function of $\alpha$ [see Figs.~\ref{fig:5}(d) and ~\ref{fig:5}(e) for $V_0=\omega$]. The primary minima  in $\Pi_2^{|gg\rangle}$ (occur when $J_0(\alpha)=0$) coincide with the secondary minima of $\Pi_2^{|ee\rangle}$ ($J_{n_2=-1}(\alpha)=0$) and vice versa. At the secondary minima of $\Pi_2^{|gg\rangle}$, the system exhibits blockade dynamics, and the same at $\Pi_2^{|ee\rangle}$, the system undergoes Rabi oscillations between the states $|+\rangle$ and $|ee\rangle$.

\begin{figure}[hbt]
\centering
\includegraphics[width= .85\columnwidth]{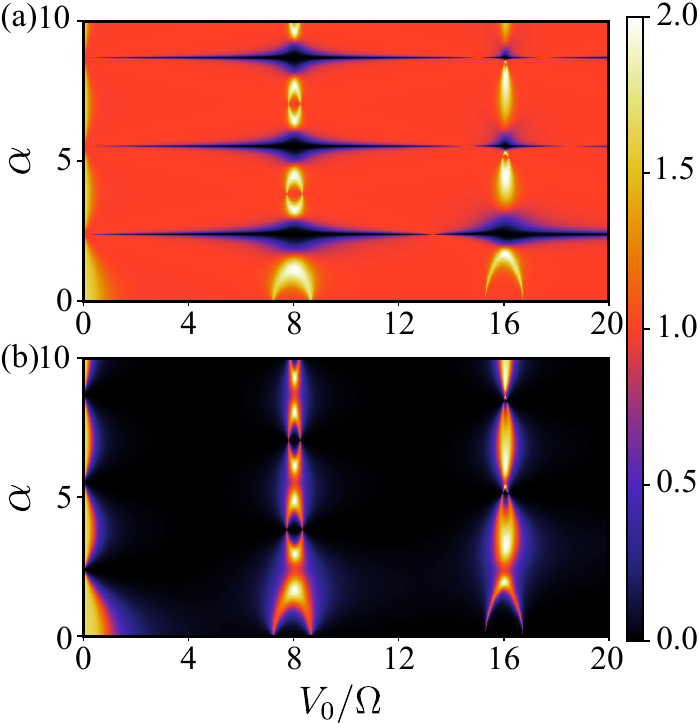} 
\caption{\small{(Color online) The IPR (a) $\Pi_2^{|gg\rangle}$ and (b) $\Pi_2^{|ee\rangle}$ as a function of $V_0$ and $\alpha$ for $N=2$, $\Delta_0=0$ (R1 resonance), and $\omega=8\Omega$. The regions of $\Pi_2^{|gg\rangle}=0$ correspond to the dynamical stabilization of $|gg\rangle$, those where both $\Pi_2^{|gg\rangle}\sim 1$ and $\Pi_2^{|ee\rangle}\sim 0$ indicate the population trapping of $|ee\rangle$, and $\Pi_2^{|gg\rangle}=2$ signals the Rydberg anti-blockade in which the system exhibits Rabi oscillations between $|gg\rangle$ and $|ee\rangle$ via the intermediate state $|+\rangle$. The intricate patterns arise due to the competition between the Rabi-couplings for the transitions $|gg\rangle\leftrightarrow|+\rangle$ [$\propto J_{n_1}(\alpha)$] and $|+\rangle\leftrightarrow|ee\rangle$ [$\propto J_{n_2}(\alpha)$]. If R2 is satisfied with $V_0=\Delta_0$ instead of R1 (a) is $\Pi_2^{|ee\rangle}$ and (b) is $\Pi_2^{|gg\rangle}$. The parameter $\alpha$ is varied by changing $\delta$ and keeping $\omega$ constant.}}
\label{fig:6} 
\end{figure}

 The maxima of both $\Pi_2^{|gg\rangle}$   and $\Pi_2^{|ee\rangle}$ in Figs.~\ref{fig:5}(d) and \ref{fig:5}(e) for $V_0=\omega$ do not coincide. At those maxima ($\Pi_2^{|gg\rangle}\sim 2$ or $\Pi_2^{|ee\rangle}\sim 2$), the system undergoes Rabi oscillations between $|gg\rangle$ and $|ee\rangle$ via the intermediate state $|+\rangle$ with an effective Rabi frequency $\propto \sqrt{J_0^2(\alpha)+J_{-1}^2(\alpha)}$. Therefore, the maxima ($\Pi_2^{|gg\rangle}=2$) in Fig.~\ref{fig:5}(d) correspond to driving-induced Rydberg anti-blockade \cite{bas18, wuj20}. Figures~\ref{fig:6}(a) and \ref{fig:6}(b) show $\Pi_2^{|gg\rangle}$ and $\Pi_2^{|ee\rangle}$, respectively for a wider range of $V_0$ and $\alpha$. In Fig.~\ref{fig:6}(a), we identify three different regions: dynamical stabilization (shown by horizontal dark regions with $\Pi_2^{|gg\rangle}\sim 0$), anti-blockade (curved shapes with $\Pi_2^{|gg\rangle}\sim 2$ around $V_0=n\omega$) and population trapping of $|ee\rangle$ ($\Pi_2^{|gg\rangle}\sim 1$ and $\Pi_2^{|ee\rangle}=0$) in the remaining majority of the parameter space. In Fig.~\ref{fig:6}(b), the dominant regions with population trapping of $|ee\rangle$ ($\Pi_2^{|ee\rangle}\sim 0$) are primarly due to the Rydberg blockade. $\Pi_2^{|ee\rangle}$ becomes non-zero in the vicinity of $V_0=n\omega$, except when $J_{n_2}\sim 0$. The non-trivial patterns in IPR we see in the $\alpha-V_0$ plane (Fig.~\ref{fig:6}) arise due to the interplay between the Rabi-couplings for the transitions $|gg\rangle\leftrightarrow|+\rangle$ [$\propto J_{n_1}(\alpha)$] and $|+\rangle\leftrightarrow|ee\rangle$ [$\propto J_{n_2}(\alpha)$].
 
{\em R2}.--- Now we analyze the population trapping of $|gg\rangle$ and $|ee\rangle$ when R2: $n_2\omega=\Delta_0-V_0$ is satisfied,  and in particular, we focus on the dynamical stabilization i.e., for $n_2=0$ or $\Delta_0=V_0$.  Following the discussions we had on R1, it is easy to see that for $V_0\ll\Omega$, the dynamical stabilization of the states $|ee\rangle$ and $|gg\rangle$ is provided by the condition, $J_0(\alpha)=0$. As $V_0$ (or equivalently $\Delta_0$) increases, the state $|gg\rangle$ completely decouples from the dynamics (except when $\Delta_0=V_0=n\omega$). In the latter case, we only have to consider the dynamical stabilization of $|ee\rangle$, which is provided again by $J_{0}(\alpha)=0$. If $\Delta_0=V_0=n\omega$, both R1 and R2 are satisfied simultaneously, the freezing of $|gg\rangle$ is provided by $J_n(\alpha)=0$ and the dynamical stabilization of $|ee\rangle$ is given by $J_{0}(\alpha)=0$. In addition, the results for R2 are identical to that of R1 with $\Delta_0=0$, $V_0=n\omega$, except that the role of $|ee\rangle$ and $|gg\rangle$ are interchanged. Therefore, Figs.~\ref{fig:6}(a) and \ref{fig:6}(b) equivalently show $\Pi_2^{|ee\rangle}$ and $\Pi_2^{|gg\rangle}$ for $V_0=\Delta_0$, respectively.

{\em R3}.--- Now, we consider the case of third resonance R3: $n_3\omega=2\Delta_0-V_0$. As mentioned earlier, the resonance condition for R3 cannot be extracted directly from the Hamiltonian in Eq.~(\ref{hro2}) or Eqs.~(\ref{c1}) and (\ref{c2}) for the Rabi couplings, and hence, they do not provide us any direct hint on how dynamical stabilization is related to the  Bessel roots. When R3 is satisfied, the system exhibits Rabi oscillations between $|gg\rangle$ and $|ee\rangle$. Note that, for $V_0\ll\Omega$, the resonances R1, R2, and R3 are not well separated, and all three states ($|gg\rangle$, $|+\rangle$, $|ee\rangle$) are relevant in the dynamics which leads to the population transfer between $|gg\rangle$ and $|ee\rangle$ via $|+\rangle$ state. For large values of $V_0$, R3 gets well isolated from R1 and R2 along the $\Delta_0$-axis. In that case, the population in $|+\rangle$ becomes negligible for sufficiently large values of $V_0/\omega$, except when $\Delta_0=n\omega$. For small values of both RRIs and $\Delta_0$ compared to the driving frequency, i.e., for  $\Delta_0/\omega\ll 1$ and $V_0/\omega\ll 1$, we obtain an effective Hamiltonian as, 
\begin{widetext}
\begin{equation}
\hat H_{{\rm eff}}^{(\Delta_0\ll\omega, V_0\ll\omega)}\simeq \frac{\Omega J_0(\alpha)}{2}\left(1-i\pi\frac{\Delta_0}{\omega}\right)\sum_{j=1}^2\hat\sigma_{eg}^j+\frac{i^{n_3}\Omega }{2}\left(J_{n_3}(\alpha)-J_0(\alpha)\right)\left(1+i\pi\frac{\Delta_0}{\omega}\right)\hat X+{\rm H.c.}.
\label{heff4}
\end{equation}
\end{widetext}

\begin{figure}
\centering
\includegraphics[width=\columnwidth]{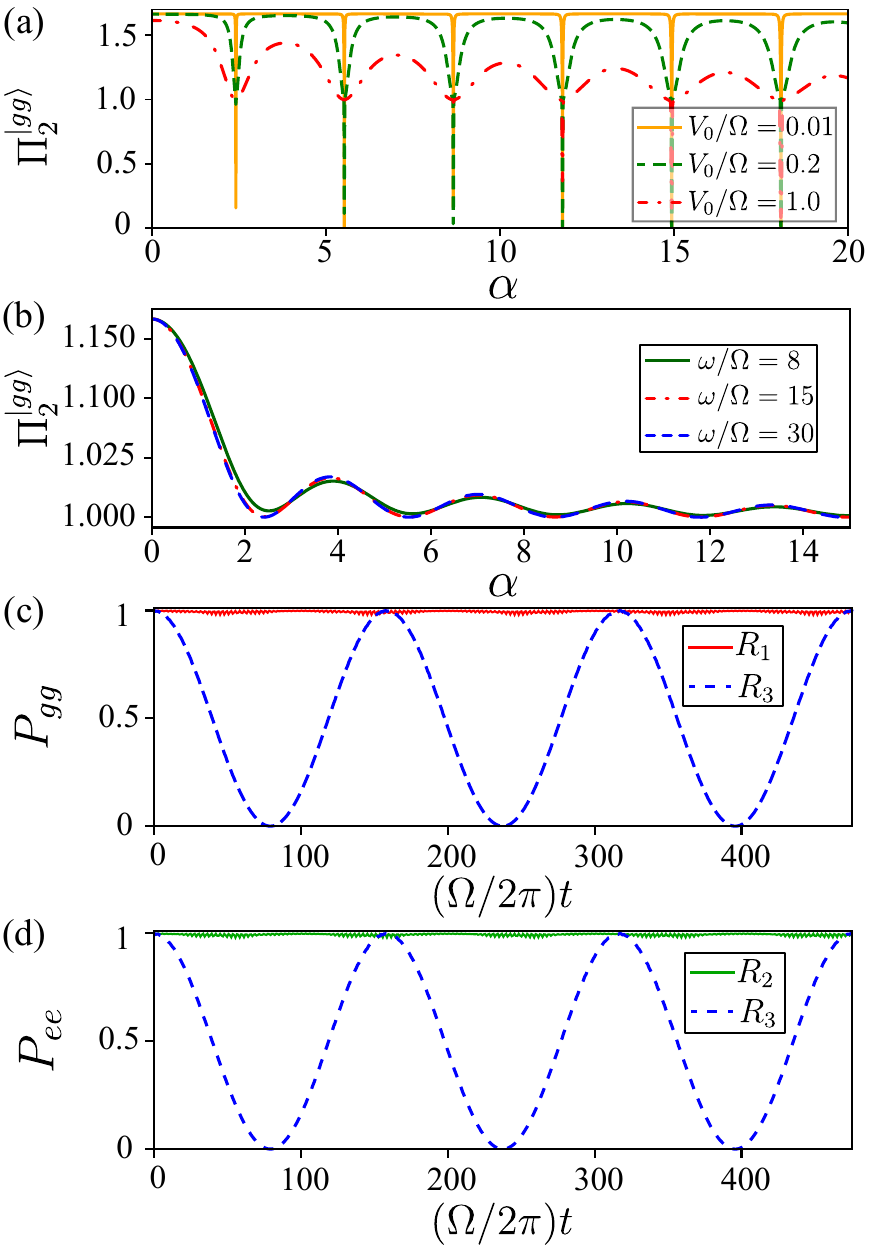} 
\caption{\small{(Color online) (a) The IPR ($\Pi_2^{|gg\rangle}$) as a function of $\alpha$ for $\omega=30\Omega$ for different $V_0$ satisfying the R3 resonance with $n_3=0$, i.e., $2\Delta_0=V_0$. (b) The same as in (a), but for different $\omega$ and $V_0=6\Omega$. In (c), we show the dynamics for the initial state $|gg\rangle$ for the two different resonances:  R1 (solid line with $n_1=0$) and R3 (dashed line with $n_3=0$) at the first root of $J_0(\alpha)$, $\omega=15\Omega$ and $V_0=6\Omega$. In (d), we show the same as in (c), except that the initial state is $|ee\rangle$ and for the resonances: R2 (solid line with $n_2=0$) and R3 (dashed line with $n_3=0$). The parameter $\alpha$ is varied by changing $\delta$ and keeping $\omega$ constant.}}
\label{fig:7} 
\end{figure} 

When $n_3=0$, the second term with $\hat X$ in Eq.~(\ref{heff4}) vanishes, and the dynamical stabilization of both $|gg\rangle$ and $|ee\rangle$ is provided by the zeros of $J_0(\alpha)$. This result has been further verified by numerical calculations of the Schr\"odinger equation, using the crossings in the Floquet spectrum and IPR [see Fig.~\ref{fig:7}(a)]. In contrast with R1 and R2, as $V_0$ increases, the dynamical stabilization for R3 demands both higher driving frequencies ($\omega$) and larger modulation amplitudes ($\alpha$).  As shown in Fig.~\ref{fig:7}(a), for $V_0=0.01\Omega$, we get the IPR almost identical to that of the non-interacting case [see Fig.~\ref{fig:1}(f)], which exhibits sharp minima  at $J_0(\alpha)=0$. For a fixed $\omega$, increasing $V_0$ makes the minima broader, and in particular, those at small values of $\alpha$ get lifted from zero. That means, increasing $V_0/\omega$ destroys dynamical stabilization at small values of $\alpha$ as seen for $V_0=0.2\Omega$ and $V_0=1\Omega$ in Fig.~\ref{fig:7}(a). In Fig.~\ref{fig:7}(b), we show IPR at a sufficiently large value of RRIs ($V_0=6\Omega$) and for different $\omega$, and we see that the sharp minima with vanishing IPR have disappeared completely and become smooth minima. These results can be understood from Eqs.~(\ref{c1}) and (\ref{c2}). For sufficiently large $V_0$, satisfying resonance condition $2\Delta_0=V_0$ does not select a single Bessel function for the Rabi couplings, which hinders the dynamical stabilization. This strong dependence of $V_0$ on the dynamical stabilization under R3 resonance, is in high contrast with that of R1 and R2. To show that explicitly, we look at the dynamics at the first Bessel zero of $J_0(\alpha)$ for the three resonances R1, R2, and R3 for sufficiently large $V_0$ [see Figs.~\ref{fig:7}(c) and \ref{fig:7}(d)]. In Fig.~\ref{fig:7}(c), we show the dynamics for the initial state $|gg\rangle$, satisfying resonances R1 and R3, and in Fig.~\ref{fig:7}(c), the dynamics is shown for the initial state $|ee\rangle$ satisfying R2 and R3. In both figures, we observe population dynamics for R3, indicating the absence of dynamical stabilization at large RRI.

\subsection{Dynamical stabilization of maximally entangled Bell states}
In the following, we consider the dynamical stabilization of two class of Bell states: $|+\rangle$ and $|B+\rangle=(|gg\rangle+|ee\rangle)/\sqrt{2}$, and they both are maximally entangled two-qubit states. We use the bipartite entanglement entropy to characterize the correlation or entanglement between the qubits. To quantify it, we label the qubits as $A$ and $B$, and the entanglement entropy of subsystem $A$ is obtained as $\mathcal S_A=-\rm{Tr}(\rho_A\log_2\rho_A)=-\sum_{k}\lambda_k\log_2\lambda_k$, where $\rho_A$ is the reduced density matrix of the subsystem $A$ and $\lambda_k$ are the eigenvalues of $\rho_A$. Both $|+\rangle$ and $|B+\rangle$ have $\mathcal S_A=1$, and under dynamical stabilization, we expect $\mathcal S_A$ also to be stabilizing over time.

\begin{figure}
\centering
\includegraphics[width= \columnwidth]{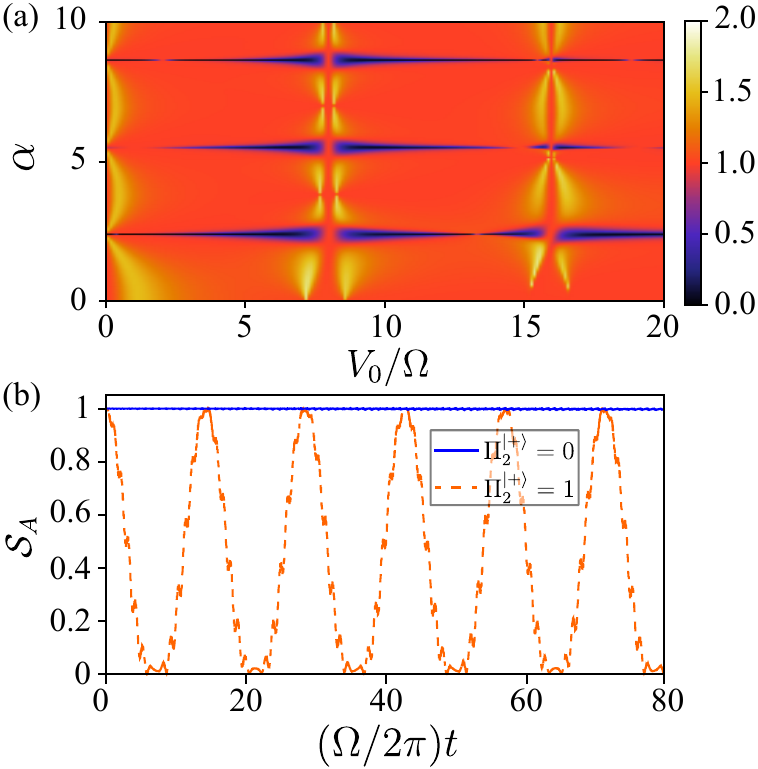} 
\caption{\small{(Color online) (a) IPR $\Pi_2^{|+\rangle}$ as a function of $\alpha$ and $V_0$ for $\omega=8\Omega$ and $\Delta_0=0$. The parameter $\alpha$ is varied by changing $\delta$ and keeping $\omega$ constant. (b) The general behavior of the dynamics of the entanglement entropy $\mathcal S_A$ for $\Pi_2^{|+\rangle}=0$ (solid line), indicating dynamical stabilization and for $\Pi_2^{|+\rangle}=1$ (dashed line).}}
\label{fig:8} 
\end{figure} 

$|+\rangle$ state.--- The state $|+\rangle$ is involved in two resonances: R1 and R2. For $V_0\ll\Omega$, the resonances R1 and R2 are not entirely separable. The latter implies that the population from $|+\rangle$ state transfers almost equally to both $|gg\rangle$ and $|ee\rangle$ states for $V_0\ll\Omega$. Following Eq.~(\ref{heff3}) for $V_0/\omega\ll 1$, we can see that dynamical stabilization of $|+\rangle$ occurs when $J_0(\alpha)=0$. For sufficiently large $V_0$ (excluding $V_0=n\omega$), the resonances R1 and R2 can be well isolated from each other, and the dynamical stabilization of $|+\rangle$ is still determined by the zeros of $J_0(\alpha)$ if either R1 or R2 is satisfied. If R1 alone is satisfied, the Rydberg blockade prevents any transition to $|ee\rangle$ thereby stabilizing $|+\rangle$ state dynamically at $J_0(\alpha)=0$.  On the other hand, the resonance condition R2 demands a large detuning, which prevents any population transfer from $|+\rangle$ to $|gg\rangle$. The latter helps the dynamical stabilization of state $|+\rangle$. Note that, when $|+\rangle$ is dynamically stabilized, one of the Floquet modes becomes $|+\rangle$, as we have discussed in Sec.~\ref{ps}.

Keeping $n_1=0$ and for $V_0=n\omega$ with $n$ being a non-zero integer, both R1 and R2 are satisfied simultaneously, and the dynamical stabilization of $|+\rangle$ requires both $J_0(\alpha)=0$ and $J_{-n}(\alpha)=0$. The latter criteria can never be satisfied with $n\neq 0$, which prevents the dynamical stabilization of $|+\rangle$. This implies that the entangled state is harder to stabilize dynamically than the product state $|gg\rangle$. The above results are summarized in Fig.~\ref{fig:8}(a), in which we show the IPR $\Pi_2^{|+\rangle}$ as a function of $\alpha$ and $V_0$.  The broken horizontal stripes  in Fig.~\ref{fig:8}(a) correspond to the regions of dynamical stabilization of  $|+\rangle$ state. The regions with $\Pi_2^{|+\rangle}=1$ correspond to the blockade dynamics and those with $\Pi_2^{|gg\rangle}=2$ indicate that all three states are very involved in the dynamics. As expected, for R2 resonance and $V_0=\Delta_0$, we get  the same results as above, with the only difference is that the regions with $\Pi_2^{|+\rangle}=1$ indicate the Rabi oscillations between $|+\rangle$ and $|ee\rangle$. Further, the time evolution of the entanglement entropy for the initial state $|+\rangle$ and different IPR is shown in Fig.~\ref{fig:8}(b). As seen in Fig.~\ref{fig:8}(b), when $\Pi_2^{|+\rangle}=0$, we hardly find any dynamics in $\mathcal S_A$, which indicates that the correlation between the two atoms is preserved under the periodic driving. For the case in which $\Pi_2^{|+\rangle}=1$, the entropy $\mathcal S_A$ undergoes periodic oscillations, and for the particular case shown in Fig.~\ref{fig:8}(b), the oscillations in $\mathcal S_A$ are due to the Rabi oscillations between the entangled state $|+\rangle$ and the product state $|gg\rangle$.

$|B+\rangle$ state.--- To discuss the dynamical stabilization of the Bell state $|B+\rangle$, we need to consider the resonances, which includes either $|gg\rangle$ or $|ee\rangle$, or both. Such resonances can drive the system out of the $|B+\rangle$ state. We comment on the case where both $|gg\rangle$ and $|ee\rangle$ are involved in the resonant dynamics. The latter happens when either R3 is satisfied or both R1 and R2 are met simultaneously. As already mentioned, when the primary resonance of R3 is met ($2\Delta_0=V_0$), the system exhibits Rabi oscillations between $|gg\rangle$ and $|ee\rangle$ via $|+\rangle$. For large $V_0$, the population in $|+\rangle$ can be neglected, and $|B+\rangle$ becomes the stationary state of the unmodulated system. Therefore, the question of dynamical stabilization is irrelevant, and periodic modulation may make $|B+\rangle$ dynamically unstable. For small RRIs and $V_0/\omega\ll 1$, the dynamical stabilization is provided by the roots of $J_0(\alpha)$, which can be easily seen from Eq.~(\ref{heff4}). On the other hand, satisfying R1 and R2 conditions simultaneously requires two different Bessel functions to vanish at the same value of $\alpha$, which is never possible, ruling out the possibility of dynamical stabilization of $|B+\rangle$.
\section{Experimental Parameters}
Finally, we comment on the experimental setup and parameters, which can be used to investigate our findings. We consider a Rydberg $n S_{1/2}$  state of a rubidium atom. The two atom setups are easily realizable in labs using either optical tweezers or optical micro traps \cite{beg13}. Moreover, the interaction strengths between the Rydberg atoms can be controlled precisely by adjusting the separation between the atoms or using external fields \cite{beg13}. As we mentioned before, the periodic modulation can be generated by applying an additional oscillating RF field, which creates sidebands in the Rydberg state as shown in \cite{mil16, bas10, tau08, yue16}. Further control over the sidebands, selecting even or odd bands, are accessible via ac or dc electric fields \cite{mil16}. An alternative way, as demonstrated in a recent experiment, an intensity-modulated off-resonance laser is used to vary the energy of the intermediate excited state sinusoidally, in a two-photon transition to the Rydberg state from the ground state \cite{cla19}. The latter approach is equivalent to modulating the effective light field, which couples the ground to the Rydberg state.

Taking a typical Rabi frequency of $\Omega=1$ MHz, our studies use interaction strengths $V_0=0-20$ MHz, and modulation frequency $\omega=0-30$ MHz. Considering the Rydberg state to be $|e\rangle\equiv|45 S_{1/2}\rangle$ of a rubidium atom, which can be coupled from the ground state $|g\rangle\equiv|5 S_{1/2}\rangle$ via a two-photon transition. As we can see, the frequency differences between neighboring states are $(E_{45S_{1/2}}-E_{44S_{1/2}})/\hbar=92.96$ GHz and $(E_{46S_{1/2}}-E_{45S_{1/2}})/\hbar=86.53$ GHz, ensures that sidebands generated by the periodic modulation do not populate the nearest Rydberg states. The latter can also be suppressed by taking moderately strong modulated field, for instance, the intensity of the oscillating RF field \cite{yue16,mil16}.

\section{Summary}
\label{sum}
In summary, we have studied the Dynamical stabilization of a set of experimentally relevant product and entangled states in a  Rydberg atom pair. The presence of Rydberg-Rydberg interactions leads to state-dependent population trapping. As we have shown, unlike in the case of a single two-level atom, the population trapping or  dynamical stabilization in two interacting Rydberg atoms may not be accompanied by level crossings in the Floquet spectrum. We have discussed the dynamical stabilization of a few selected states, including both product and entangled Bell states. The latter case offers a way to preserve entanglement or correlation between two qubits for sufficiently long times, with limitations arising only from the decoherent processes. Our analysis reveals that the driving parameters are more restricted to stabilize the entangled states compared to the product states dynamically. The results we have discussed here on population trapping or dynamical stabilization are valid for a pair of any interacting two-level systems. 

Our studies immediately raise the question of population trapping or dynamical stabilization in extended systems, i.e., beyond a pair of atoms. For instance, it would be interesting to analyze how the population trapping affects the bipartite and tripartite entanglement of W and GHZ-states in three or more atoms setup. As the number of qubits or atoms increases, the Floquet spectrum's complexity also increases, making the scenario more intriguing. 

\section{Acknowledgments}
We acknowledge UKIERI-UGC Thematic Partnership No. IND/CONT/G/16-17/73 UKIERI-UGC project, UGC for UGC-CSIR NET-JRF/SRF, and the support from the EPSRC through Grant No. EP/R04340X/1 via the QuantERA project ERyQSenS. R.N. further acknowledges DST-SERB for Swarnajayanti fellowship File No. SB/SJF/2020-21/19.

\appendix
\section{Derivation of Eq.~(\ref{heff3})}
\label{a1}
When R1 resonance ($n_1\omega=\Delta_0$) is satisfied, we can write Eq.~(\ref{hro2}) as,
\begin{eqnarray}
 \hat H'=\frac{\Omega}{2}i^{n_1} J_{n_1}(\alpha)\left(\sum_{j=1}^2\hat\sigma_{eg}^j+\hat X\left(e^{iV_0t}-1\right)\right)+\nonumber \\
 \frac{\Omega}{2}\sum_{m\neq n_1}i^m J_m(\alpha)e^{i(m-n_1)\omega t}\left(\sum_{j=1}^2\hat\sigma_{eg}^j+\hat X\left(e^{iV_0t}-1\right)\right)+ {\rm H.c.}, \nonumber\\
 \label{ha1}
 \end{eqnarray}
where the first term provides us the resonant contribution. In the limit $\omega\gg\Omega$ and $\omega\gg V_0$ the contribution from the second term in Eq.~(\ref{ha1}) is negligible and we can obtain an effective time independent Hamiltonian $H_{{\rm eff}}=1/T\int_0^Tdt \ \hat H'(t)$ where $T=2\pi/\omega$ as,
\begin{eqnarray}
 \hat H_{{\rm eff}}=\frac{\Omega}{2}i^{n_1} J_{n_1}(\alpha)\left(\sum_{j=1}^2\hat\sigma_{eg}^j+\hat X\left(\frac{e^{iV_0T}-1}{iV_0T}-1\right)\right)+\nonumber \\
 \frac{\Omega}{2}\sum_{m\neq n_1}i^m J_m(\alpha)\hat X\left(\frac{e^{iV_0T}-1}{i[(m-n_1)\omega+V_0]T}\right)+ {\rm H.c.}, \nonumber\\
 \label{ha1}
 \end{eqnarray}
 In leading orders of $V_0/\omega$, we get the Eq.~(\ref{heff3}) in the main text. In a similar manner, we can derive the effective Hamiltonian in Eq.~(\ref{heff4}).
\bibliographystyle{apsrev4-1}
\bibliography{lib.bib}
\end{document}